\documentclass{stacs_proc}

\usepackage{amsmath}
\usepackage{amsfonts}
\usepackage{epsfig}
\usepackage{graphics}

\newcommand{\field}[1]{\mathbb{#1}}
\newcommand{\Q}{\field{Q}}
\def\pp{\mathinner{\ldotp\ldotp}}   

\begin{document}

\title{Understanding maximal repetitions in strings}

\author[ref1]{M. Crochemore}{Maxime Crochemore}
\address[ref1]{King's College London, Strand, London WC2R 2LS, United 
Kingdom
\newline and Institut Gaspard-Monge, Universit\'e Paris-Est, France}
\email{maxime.crochemore@kcl.ac.uk}

\author[ref2]{L. Ilie}{Lucian Ilie}
\address[ref2]{Department of Computer Science, University of Western Ontario
\newline N6A 5B7, London, Ontario, Canada}
\email{ilie@csd.uwo.ca}

\thanks{This work has been done during the second author's stay at
Institut Gaspard-Monge.  The same author's research was supported in
part by NSERC}




\begin{abstract}
    The cornerstone of any algorithm computing all repetitions in a
    string of length $n$ in ${\mathcal O}(n)$ time is the fact that
    the number of runs (or maximal repetitions) is ${\mathcal O}(n)$.
    We give a simple proof of this result.  As a consequence of our
    approach, the stronger result concerning the linearity of the sum
    of exponents of all runs follows easily.
\end{abstract}

\maketitle

\stacsheading{2008}{11-16}{Bordeaux}
\firstpageno{11}

\section{Introduction}

Repetitions in strings constitute one of the most fundamental areas
of string combinatorics with very important applications to text
algorithms, data compression, or analysis of biological sequences.
One of the most important problems in this area was finding an
algorithm for computing all repetitions in linear time.
A major obstacle was encoding all repetitions in linear space
because there can be $\Theta(n\log n)$ occurrences of squares in a
string of length $n$ (see \cite{Cr81}).
All repetitions are encoded in runs (that is, maximal repetitions)
and Main~\cite{Ma89} used the s-factorization of
Crochemore~\cite{Cr81} to give a linear-time algorithm for finding
all leftmost occurrences of runs.
What was essentially missing to have a linear-time algorithm for
computing all repetitions, was proving that there are at most
linearly many runs in a string.
Iliopoulos et al.~\cite{IMS97} showed that this property is true for
Fibonacci words.
The general result was achieved by Kolpakov and
Kucherov~\cite{KoKu00} who gave a linear-time algorithm for
locating all runs in \cite{KoKu99}.

Kolpakov and Kucherov proved that the number of runs in a string
 of length $n$ is at most $cn$ but could not provide any value
 for the constant $c$.
Recently, Rytter~\cite{Ry06} proved that $c\le 5$.
The conjecture in \cite{KoKu00} is that $c=1$ for binary alphabets,
 as supported by computations for string lengths up to 31.
Using the technique of this note, we have proved \cite{CI07jcss}
 that it is smaller than $1.6$, which is the best value so far.

Both proofs in \cite{KoKu99} and \cite{Ry06} are very intricate
 and our contribution is a simple proof of the linearity.
On the one hand, the search for a simple proof is motivated by the
 very importance of the result -- this is the core of the analysis of
 any optimal algorithm computing all repetitions in strings.
None of the above-mentioned proofs can be included in a textbook.
We believe that the simple proof shows very clearly why the number
 of runs is linear.
On the other hand, a better understanding of the structure of runs
 could pave the way for simpler linear-time algorithms for finding
 all repetitions.
For the algorithm of \cite{KoKu99} (and \cite{Ma89}), relatively
 complicated and space-consuming  data structures are needed,
 such as suffix trees.

The technical contribution of the paper is based on the notion of
 $\delta$-close runs (runs having close centers), which is an improvement
 on the notion of neighbors (runs having close starting positions)
 introduced by Rytter \cite{Ry06}.

On top of that, our approach enables us to derive easily the stronger
 result concerning the linearity of the sum of exponents of all runs
 of a string.
Clearly this result implies the first one, but the converse is
 not obvious.
The second result was given another long proof in \cite{KoKu00};
 it follows also from \cite{Ry06}.

Finally, we strongly believe that our ideas in this paper can be
 further refined to improve significantly the upper bound on the number of
 runs, if not to prove the conjecture.
The latest refinements and computations (December 2007)
 show a $1.084n$ bound.

\section{Definitions}

Let $A$ be an alphabet and $A^*$ the set of all finite strings over $A$.
We denote by $|w|$ the length of a string $w$, by $w[i]$ its $i$th letter,
 and by $w[i\pp j]$ its factor $w[i]w[i+1]\cdots w[j]$. We say
that $w$ has period $p$ iff $w[i]=w[i+p]$, for all $1\le i \le
|w|-p$.
The smallest period of $w$ is called {\it the period} of $w$ and the
ratio between the length and the period of $w$ is called the {\it exponent}
 of $w$.

For a positive integer $n$, the $n$th power of $w$ is defined
inductively by $w^1=w$, $w^n = w^{n-1}w$.
A string is {\it primitive} if it cannot be written as a proper
 integer (two or more) power of another string.
Any nonempty string can be uniquely written as an integer power of a
primitive string, called its {\it primitive root}.
It can also be uniquely written in the form $u^ev$ where $|u|$ is its
 (smallest) period, $e$ is the integral part of its exponent,
 and $v$ is a proper prefix of $u$.

The following well-known {\it synchronization} property will be
useful: If $w$ is primitive, then $w$ appears as a factor of $ww$
only as a prefix and as a suffix (not in-between).
Another property we use is {\it Fine and Wilf's periodicity lemma}:
If $w$ has periods $p$ and $q$ and $|w|\ge p+q$, then $w$ has also
period $\gcd(p,q)$.
(This is a bit weaker than the original lemma which works as soon as
$|w|\ge p+q-\gcd(p,q)$, but it is good enough for our purpose.)
We refer the reader to \cite{Lo02} for all concepts used here.

For a string $w=w[1\pp n]$, a {\it run}%
\footnote{Runs were introduced in \cite{Ma89} under the name
{\it maximal periodicities}; the are called {\it m-repetitions}
 in \cite{KoKu00} and {\it runs} in \cite{IMS97}.}
 (or maximal repetition) is an interval $[i\pp j]$, $1\leq i<j\leq n$,
 such that (i) the factor $w[i\pp j]$ is periodic
 (its exponent is 2 at least)
 and (ii) both $w[i-1\pp j]$ and $w[i\pp j+1]$,
 if defined, have a strictly higher (smallest) period.
As an example, consider $w = {\sf abbababbaba}$; $[3\pp 7]$ is a run
 with period 2 and exponent 2.5;
 we have $w[3\pp 7] = {\sf babab} = ({\sf ba})^{2.5}$.
Other runs are $[2\pp 3], [7\pp 8], [8\pp 11], [5\pp 10]$ and $[1\pp 11]$.
For a run starting at $i$ and having period $|x|=p$, we shall call
 $w[i\pp i+2p-1]=x^2$ the {\it square} of the run (this is the only
 part of a run we can count on).
Note that $x$ is primitive and the square of a run cannot
 be extended to the left (with the same period) but may be
 extendable to the right.
The {\it center} of  the run is the position $c=i+p$.
We shall denote the {\it beginning} of the run
 by $i_x = i$, the {\it end of its square} by $e_x=i_x+2p-1$,
 and its {\it center} by $c_x=i_x+p$.

\section{Linear number of runs}

We describe in this section our proof of the linear number of runs.
The idea is to partition the runs by grouping together those
having close centers and similar periods.
To this aim, for any $\delta>0$, we say that two runs having squares
 $x^2$ and $y^2$ are {\it $\delta$-close} if (i) $|c_x-c_y| \le \delta$
 and (ii) $2\delta \le |x|,|y| \le 3\delta$.
We prove that there cannot be more than three mutually $\delta$-close runs.
(There is one exception to this rule -- case (vi) below -- but then,
 even fewer runs are obtained.)
This means that the number of runs with the periods between $2\delta$ and
$3\delta$ in a string of length $n$ is at most $\frac{3n}{\delta}$. Summing up
for values $\delta_i=\frac{1}{2}\bigl(\frac{3}{2}\bigr)^i$, $i\ge 0$,
all periods are considered and we obtain that the number of runs is at most
\begin{equation}
\label{eq_linear}
\sum_{i=0}^{\infty} \frac{3n}{\delta_i}
= \sum_{i=0}^{\infty} \frac{3n}{\frac{1}{2}(\frac{3}{2})^i}
 = 18n.
\end{equation}

For this purpose, we start investigating what happens when three
 runs in a string $w$ are $\delta$-close.
Let us denote their squares by $x^2, y^2, z^2$, their periods by
 $|x|=p$, $|y|=q$, $|z|=r$, and assume $p\le q\le r$.
We discuss below all the ways in which $x^2$ and $y^2$ can be
 positioned relative to each other and see that long factors of both
 runs have small periods which $z^2$ has to synchronize.
This will restrict the beginning of $z^2$ to only
 one choice as otherwise some run would be left extendable.
Then a fourth run $\delta$-close to the previous three cannot exist.

Notice that, for cases (i)-(v) we assume the centers of the runs are different;
the case when they coincide is covered by (vi).

\begin{figure}[t]
\begin{center}
\includegraphics[width=\textwidth]{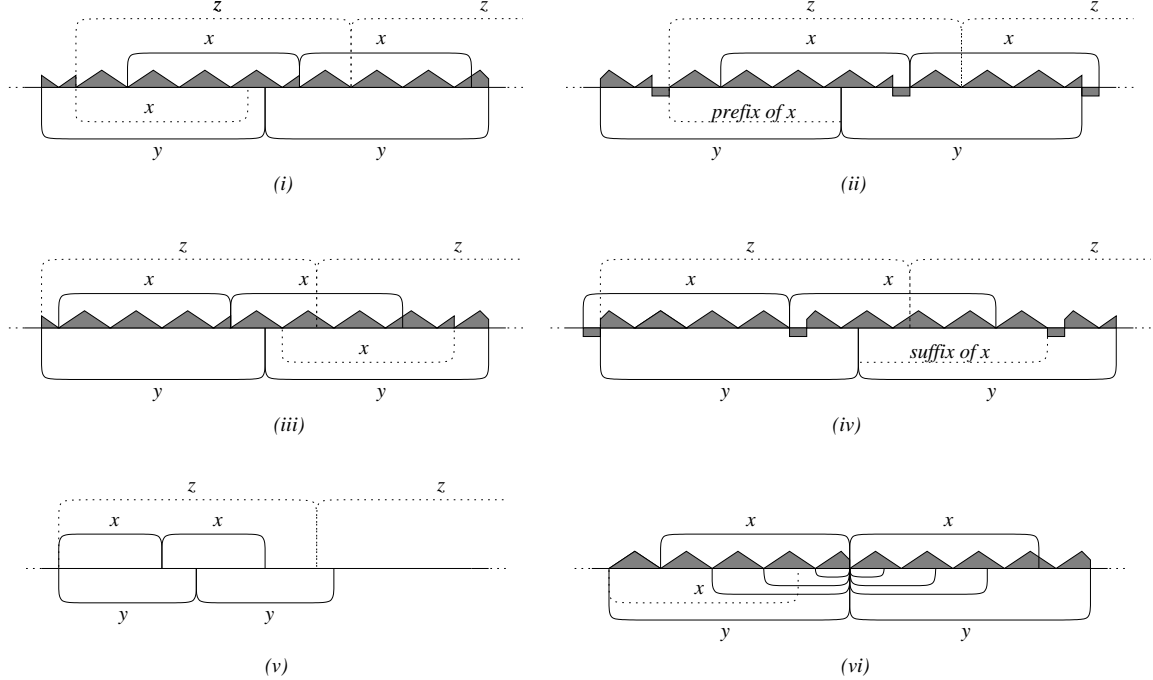}
\caption{Relative position of $x^2$ and $y^2$.}
\label{fig1}
\end{center}
\end{figure}

(i) $(i_y < i_x <) c_y < c_x < e_x \le e_y$.
Then $x$ and the suffix of length $e_y-c_x$ of $y$
have period $q-p$; see Fig.~\ref{fig1}(i).
We may assume the string corresponding to this period is a primitive
 string as otherwise we can make the same reasoning with its
 primitive root.

Since $z^2$ is $\delta$-close to both $x^2$ and $y^2$, it must be that
$c_z\in [c_x-\delta \pp  c_y+\delta]$. Consider the interval of length $q-p$
that ends at the leftmost possible position for $c_z$, that is,
$I=[c_x-\delta-(q-p) \pp  c_x-\delta-1]$. It is included in the first period of $z^2$,
that is, $[i_z \pp  c_z-1]$, and in $[i_x \pp  c_y]$.
Thus  $w[I]$ is primitive and equal, due to $z^2$, to $w[I+r]$ which is a factor of
$w[c_x \pp  e_y]$. Therefore, the periods inside the former must synchronize
with the ones in the latter. It follows, in the case $i_z>i_x-(q-p)$, that
$w[i_z-1]=w[c_z-1]$, that is, $z^2$ is left extendable, a contradiction.
If $i_z<i_x-(q-p)$, then $w[c_x-1]=w[i_x-(q-p)-1]=w[i_x-1]$, that is, $x^2$
is left extendable, a contradiction. The only possibility is that $i_z=i_x-(q-p)$
and $r$ equals $q$ plus a multiple of $q-p$.
Here is an example: $w = {\sf baabababaababababaab}$,
 $x^2=w[5\pp 14]=({\sf ababa})^2$, $y^2=w[1\pp 14]=({\sf baababa})^2$,
 and $z^2=w[3\pp 20]=({\sf abababaab})^2$.

We have already, due to $z^2$, that $x=\rho^{\ell}\rho'$, where $|\rho| = q-p$
and $\rho'$ a prefix of $\rho$. A fourth run $\delta$-close to the previous three
would have to have the same beginning as $z^2$ and the length of its period
would have to be also $q$ plus a multiple of $q-p$. This would imply
an equation of the form $\rho^m\rho'=\rho'\rho^m$ and then $\rho$ and $\rho'$
are powers of the same string, a contradiction with the primitivity of $x$.

(ii) $(i_y < i_x <) c_y < c_x < e_y \le e_x$; this is similar with (i);
see Fig.~\ref{fig1}(ii). Here the prefix of length $e_y-c_x$ of $x$ is a suffix
of $y$ and has period $q-p$.

(iii) $i_y < i_x < c_x < c_y (< e_x < e_y)$. Here $x$ and the prefix of
length $c_x-i_y$ of $y$ have period $q-p$; see Fig.~\ref{fig1}(iii).
As above, a third $\delta$-close run $z^2$ would have to share the
 same beginning with $y^2$, otherwise one of $y^2$ or $z^2$ would be
 left extendable.
A fourth $\delta$-close run would have to start at the same place
 and, because of the three-prefix-square lemma%
\footnote{For three words $u,v,w$, it states that if $uu$ is a prefix of $vv$,
 $vv$ is a prefix of $ww$, and $u$ is primitive, then $|u|+|v|\leq |w|$.}
 of \cite{CrRy95}, since $p$
 is primitive, it would have a period at least $q+r$, which is
 impossible.

(iv) $i_x < i_y(< c_x < c_y < e_x < e_y)$; this is similar with (iii);
see Fig.~\ref{fig1}(iv).
A third run would begin at the same position as $y^2$ and there is
 no fourth run.

(v) $i_x=i_y$; see Fig.~\ref{fig1}(v). Here not even a third $\delta$-close
run exists because of the three-square lemma that implies $r\ge p+q$.

(vi) $c_x=c_y$.
This case is significantly different from the other ones, as we can have
 many $\delta$-close runs here. However, the existence of many runs with the same center
 implies very strong periodicity properties of the string which allow us to
 count the runs globally and obtain even fewer runs than before.

In this case both $x$ and $y$ have the same small period $\ell=q-p$;
 see Fig.~\ref{fig1}(vi).
If we note $c=c_y$ then we have $h$ runs $x^{\alpha_j}_j$, $1\le j \le h$,
 beginning at positions $i_{x_j}=c-((j-1)\ell+\ell')$, where $\ell'$ is the
 length of the suffix of $x$ that is a prefix of the period.

We show that in this case we have less runs than as counted in the sum (\ref{eq_linear}).
For $h\le 9$ there is nothing to prove as no four of our $x_j^{\alpha_j}$ runs are
counted for the same $\delta$. Assume $h\ge 10$. There exists $\delta_i$
such that $\frac{\ell}{2} \le \delta_i \le \frac{3\ell}{4}$, that is, this
$\delta_i$ is considered in (\ref{eq_linear}). Then it is not difficult to see
that there is no run in $w$ with period between $\ell$ and $\frac{9}{4}\ell$
and center inside $J=[c+\ell+1 \pp  c+(h-2)\ell+\ell']$. But
$\ell \le 2\delta_i < 3\delta_i \le \frac{9}{4}\ell$ and the length of $J$
is $(h-3)\ell+\ell' \ge (h+1)\delta_i$.
This means that at least $h$ intervals of length $\delta_i$
in the sum (\ref{eq_linear}) are covered by $J$ and therefore
at least $3h$ runs in (\ref{eq_linear}) are replaced by our $h$ runs.

We need also mention that these $h$ intervals of length $\delta_i$ are not
reused by a different center with multiple runs since such centers
cannot be close to each other. Indeed, if we have two centers $c_j$ with the above
parameters $h_j, \ell_j$, $j=1,2$, then, as soon as the longest runs overlap
over $\ell_1+\ell_2$ positions, we have $\ell_1=\ell_2$, due to Fine and Wilf's
lemma. Then, the closest positions of
$J_1$ and $J_2$ cannot be closer than $\ell_1=\ell_2\ge\delta_i$ as this would
make some of the runs non-primitive, a contradiction.
Thus the bound in (\ref{eq_linear}) still holds and we proved

\begin{theorem}
The number of runs in a string of length $n$ is $\mathcal{O}(n)$.
\end{theorem}

\section{The sum of exponents}

Using the above approach, we show in this section that the sum of exponents of all runs is also
linear. The idea is to prove that the sum of exponents of all runs with the centers
in an interval of length $\delta$ and periods between $2\delta$ and $3\delta$  is
less than 8. (As in the previous proof, there are exceptions to this rule, but in those
cases we get a smaller sum of exponents.)
Then a computation similar to (\ref{eq_linear}) gives that
the sum of exponents is at most $48n$.

To start with, Fine and Wilf's periodicity lemma can be rephrased as follows:
For two primitive strings $x$ and $y$, any powers $x^{\alpha}$ and
$y^{\beta}$ cannot have a common factor longer than $|x|+|y|$ as such a factor would
have also period $\gcd(|x|,|y|)$, contradicting the primitivity of $x$ and $y$.

Next consider two $\delta$-close runs, $x^{\alpha}$ and $y^{\beta}$,
$\alpha,\beta\in\Q$. It cannot be that both $\alpha$ and $\beta$ are
$2.5$ or larger, as this would imply an overlap of length at least $|x|+|y|$ between the
two runs, which is forbidden by Fine and Wilf's lemma since $x$ and $y$ are primitive.
Therefore, in case we have three mutually $\delta$-close runs,
two of them must have their exponents smaller than $2.5$.
If the exponent of the third run is less than 3, we obtain the total of 8 we were
looking for. However, the third run, say $z^{\gamma}$, $\gamma\in\Q$,
 may have a larger exponent.
If it does, that affects the runs in the neighboring intervals of length $\delta$.
More precisely, if $\gamma\ge 3$, then there cannot be any center of run with
period between $2\delta$ and $3\delta$ in the next (to the right) interval of
length $\delta$.
Indeed, the overlap between any such run and $z^{\gamma}$ would imply,
as above, that their roots are not primitive, a contradiction.
In general, the following $\lfloor 2(\gamma-2.5)\rfloor$ intervals of
length $\delta$ cannot contain any center of such runs.
Thus, we obtain a smaller sum of exponents when this situation is met.

The second exception is given by case (vi) in the previous proof, that is,
when many runs share the same center; we use the same notation as in (vi).
We need to be aware of the exponent of the run $x_1^{\alpha_1}$, with the smallest period,
as $\alpha_1$ can be as large
as $\ell$ (and unrelated to $h$, the number of runs with the same center).
We shall count $\alpha_1$ into the appropriate interval of length $\delta_i$;
notice that $x_1^{\alpha_1}$ and $x_2^{\alpha_2}$ are never $\delta$-close,
for any $\delta$, because $|x_2|>2|x_1|$.
For $2\le j\le h-1$, the period $|x_j|$ cannot be extended by
more than $\ell$ positions to the right past the end of the initial square, and
thus $\alpha_j \le 2+\frac{1}{j}$. Therefore, their contribution to the sum of
exponents is less than $3(h-2)$. They replace the exponents of the runs
with centers in the interval $J$ and periods between $\ell$ and $\frac{9}{4}\ell$
which otherwise would contribute at least $6h$ to the sum of exponents.
The run with the longest period, $x_h^{\alpha_h}$, can have an arbitrarily high
exponent but the replaced runs in $J$ need to account only for a fraction
(3 units) of it since $\alpha_h\ge 3$ implies new centers with multiple runs
and hence new $J$ intervals (precisely $\lfloor\alpha_h-2\rfloor$)
that account for the rest. We proved

\begin{theorem}
    The sum of exponents of the runs in a string of length $n$ is
    $\mathcal{O}(n)$.
\end{theorem}

\end{document}